\documentclass[sigplan,screen]{acmart}
\AtBeginDocument{%
  \providecommand\BibTeX{{%
    \normalfont B\kern-0.5em{\scshape i\kern-0.25em b}\kern-0.8em\TeX}}}
    
\usepackage{hyperref}

\usepackage{mathtools}
\usepackage[utf8]{inputenc}
\usepackage{algpseudocode}
\usepackage{algorithm}

\usepackage{caption}
\usepackage{subcaption}
\usepackage{multirow}
\usepackage[english=american]{csquotes} 
\setcopyright{acmcopyright}
\copyrightyear{2022}
\acmYear{2022}
\acmDOI{XXXXXXX.XXXXXXX}

\acmConference[HotCloudPerf 2022]{5th Workshop on Hot Topics in Cloud Computing Performance}{April 09--10, 2022}{Virtual Conference}
\acmPrice{15.00}
\acmISBN{978-1-4503-XXXX-X/18/06}

\usepackage{draftwatermark}
\SetWatermarkText{DRAFT}
\SetWatermarkScale{1}

\begin{document}

\title[Beauty and the beast: A Case Study on Perf. of Data-Intensive Containerized Cloud Apps]{Beauty and the beast: \\ A case study on performance prototyping of data-intensive containerized cloud applications}


\author{Floriment Klinaku}
\affiliation{%
  \institution{University of Stuttgart}
  \city{Stuttgart}
  \country{Germany}}
\email{klinaku@iste.uni-stuttgart.de}
\orcid{0000-0002-4760-5889}

\author{Martina Rapp}
\orcid{0000-0002-2703-0440}
\author{Jörg Henss}
\orcid{0000-0002-4527-211X}
\affiliation{%
  \institution{FZI Forschungszentrum Informatik}
  \city{Karlsruhe}
  \country{Germany}}
\email{{rapp, henss}@fzi.de}

\author{Stephan Rhode}
\affiliation{%
  \institution{Robert Bosch GmbH}
  \city{Renningen}
  \country{Germany}}
\email{stephan.rhode@de.bosch.com}
\orcid{0000-0002-3662-0426}

\renewcommand{\shortauthors}{Klinaku and Rapp, et al.}

\begin{abstract}
Data-intensive container-based cloud applications have become popular with the increased use cases in the Internet of Things domain. Challenges arise when engineering such applications to meet quality requirements, both classical ones like performance and emerging ones like elasticity and resilience. There is a lack of reference use cases, applications, and experiences when prototyping such applications that could benefit the research community. Moreover, it is hard to generate realistic and reliable workloads that exercise the resources according to a specification. Hence, designing reference applications that would exhibit similar performance behavior in such environments is hard. In this paper, we present a work in progress towards a reference use case and application for data-intensive containerized cloud applications having an industrial motivation. Moreover, to generate reliable CPU workloads we make use of ProtoCom, a well-known library for the generation of resource demands, and report the performance under various quality requirements in a Kubernetes cluster of moderate size. Finally, we present the scalability of the current solution assuming a particular autoscaling policy. Results of the calibration show high variability of the ProtoCom library when executed in a cloud environment. We observe a moderate association between the occupancy of node and the relative variability of execution time.

\end{abstract}





\maketitle

\section{Introduction}
Containers \cite{rosen2014linux} have become the de-facto standard for packaging and deploying microservice-based applications in the cloud. A particular class of such applications continuously processes streams of data generated by a variety of connected data-sources. The performance of such applications is business-critical. In addition to performance, elasticity and resilience have become two required quality attributes to cost-efficiently handle disruptive events like unexpected failures or changes in the demand.   

Scaling non-trivial microservice-based applications remains a challenge for service providers due to the uncertain cloud environment. 
Achieving elasticity and resilience through an upfront engineering process requires suitable prediction models. 
There is, however, a lack of reference use cases, applications and experiences matching the characteristics of data-intensive cloud applications that would allow researchers to evaluate their approaches. Two prominent reference applications are proposed to foster research of microservice-based cloud applications: TeaStore~\cite{von2018teastore} and SockShop~\cite{sockshop}. Both serve more traditional use cases of classical human-centered request-reply applications. They lack, however, a processing pipeline of continuous data and also do not use asynchronous messaging communication which is very popular in such use cases. 
In addition, when prototyping such systems, it is hard to generate realistic workloads that utilize the resources according to a given specification (e.g., the time, that operations should consume the CPU). 

To tackle the aforementioned problems we present a work in progress towards a reference use case and application for enabling research for data-intensive containerized cloud applications. To generate more reliable CPU workloads and to make the application more predictable in terms of performance, we make use of ProtoCom \cite{10.1007/978-3-540-69814-2_7}, a library for calibrating and generating CPU demands on various hardware. In addition we present two different scaling strategies for the defined application and present scalability experiments to obtain a first assessment of the capabilities of the application. 
Moreover, we investigate on the high variability of the load generation approach as our initial results showed high deviations from the expected response times. 

The focus of this work is twofold: first, in Section~\ref{sec:example}, we introduce the reference use case and present performance requirements and the chosen implementation stack; second, in Section~\ref{sec:performance}, we show the variability of the resource demand generation with ProtoCom in a cloud environment, initial scaling variants for the use case and experiments that show the scalability aspect of our microservice-based implementation. In addition, Section~\ref{sec:related_work} presents related work and finally,  Section~\ref{sec:conclusions} concludes the paper and provides an outlook on future work.


\section{Related Work} \label{sec:related_work}

Our related work can be divided into two different categories. 
On the one hand, several benchmarks have been developed as a reference for cloud applications and their performance. 
On the other hand microbenchmarking has been applied to measure the impact of virtualized environments and to quantify the performance isolation available in those.
While we can not solve these inherent problems, developers must be aware of those effects affecting performance and scalability in virtualized and containerized applications.

In \cite{nikounia2015hypervisor} Nikounia et al. introduce the noisy neighbour problem, an effect that can be observed in shared infrastructures where the activity on a neighboring core may lead to performance degradation.
They report on performance degradation of up to 16x slowdown in virtualized environments.
This is caused by noisy neighbour VMs, overcommitment and hypervisor noise.

In \cite{laaberMicrobenchmarking} Laaber et al. present their findings on using microbenchmarking to assess the performance impact in virtualized environments.
The authors performed several experiments on systems deployed in public cloud environments and report on slowdown effects ranging from 0.003\% to > 100\%.
They state that several repetitions on several VM instances are required to get robust results for microbenchmarks and to detect potential slowdowns. 

In \cite{lehrig2011performance} Lehrig et al. present experiments conducted with the ProtoCom library in a virtualized cloud environment. They show that ProtoCom is well suited to emulate CPU demands realistically when using a calibration based approach.

In addition to many benchmark and test application like Spring PetClinic~\cite{springpetclinic} or ACME Air~\cite{acmeAir}, several academic case-study systems have been developed in the past for evaluating the performance of cloud and containerised systems:
In \cite{von2018teastore} the Tea Store case-study is presented, a system for studying the performance of microservices based systems.
A similar microservices based benchmark using container technologies, the Sock Shop, is presented in \cite{sockshop}.
The CloudStore application~\cite{lehrig2018cloudstore} is a reference application for comparing different cloud providers, cloud service architectures, and assess cloud deployment options. 
The TrainTicket benchmark~\cite{DBLP:journals/tse/ZhouPXSJLD21} in addition focuses on the fault-analysis of microservice-based applications.
All four case-study systems have in common that mostly request-response semantics is employed.
Asynchronous data-centric communication patterns as typical found in IoT systems are missing.


\section{Running Example} \label{sec:example}

Before going into the implementation details of the running example in Section~\ref{sec:demonstrator}, we will explain the considered use case, its purpose and architecture in Section~\ref{sec:use_case}, and discuss performance challenges in Section~\ref{sec:challenges}.

\subsection{Reference Use Case}
\label{sec:use_case}

The herein considered use case -- called remote measuring -- is a fraction of one service package from the Bosch mobility cloud \cite{bosch21}. The mobility cloud is a cloud-based integration platform for developing and updating vehicle software and services. The services are grouped into three packages: over-the-air services (update, function call, essentials, vehicle data), data services (data integration, navigation, broker), and core services (service integration, application run time, application marketplace). Remote measuring is one service from the over-the-air vehicle data package. We consider parts of remote measuring in the implemented running example and its model representation.

The app remote measuring is designed for vehicle data acquisition campaigns. Such cloud-based campaigns are beneficial in vehicle development, fleet observation (e.g. tracking of failure codes in a delivery car fleet), predictive diagnostics, and optimization of spare part logistics in aftermarket business. Imagine a vehicle homologation task \cite{Lutz2017}, where several vehicles must collect data from test rides. Cloud-based remote measuring allows test engineers to configure and conduct a signal measurement setup through web services. Such a setup contains the number and kind of signals, their recording frequency and recording triggers. The test engineer starts the measurement through a web front end. Data is cached in the car and pushed to the cloud, where the data is stored in a database. Then, the data is converted and presented as dashboard, or exported via an API for external applications. This is more flexible and convenient in contrast to conventional workflows, where each vehicle was equipped with a signal recorder, a laptop, and a test engineer who configured the campaign, stored the data and fed it into a data center afterwards.

\begin{figure}
  \centering
  \includegraphics[width=\columnwidth]{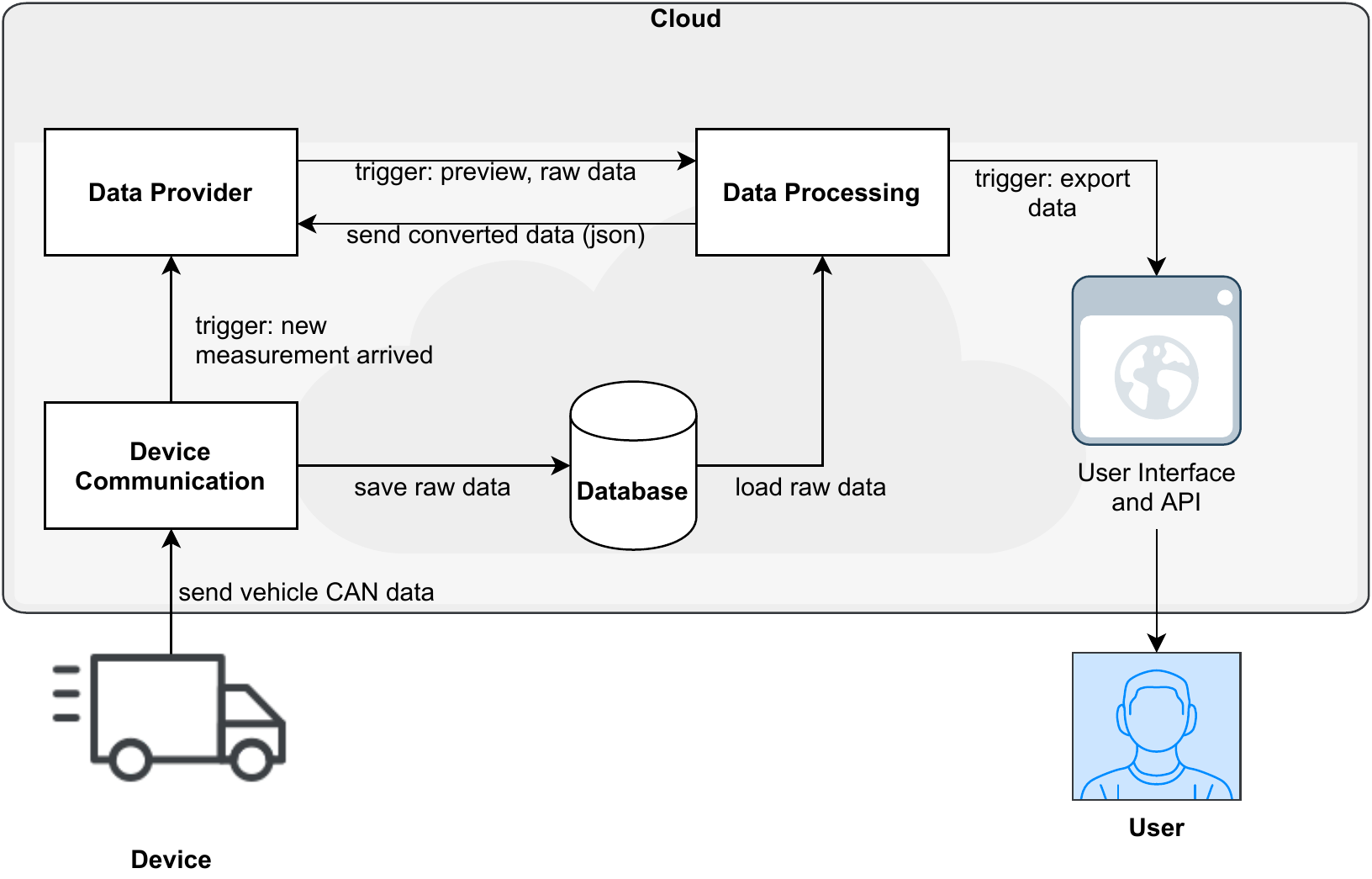}
  \caption{Architectural snipped of remote measuring application in mobility cloud suite.}
  \label{fig:application}
\end{figure}

Figure~\ref{fig:application} explains the architecture of the running example remote measuring. Starting in the lower left corner, one or more devices (vehicles) send data from their vehicle bus (CAN bus~\cite{10.5555/2636651}) to the device communication service. This service tracks the readiness of the device, caches data, stores the raw data in a data base, and triggers the next service data provider. Once, a new batch of data arrived, the data provider service triggers the data processing service, which conducts data pre-processing and data compression. For this, the data processing fetches the raw data from data base. This part of the use case is considered and implemented in the running example. 

The user interface and API on the right in Figure~\ref{fig:application} are not implemented in the running example but shown here to understand the use case from end to end. The data processing triggers an export service, which provides the converted data through API and triggers a data dashboard. The API can be used for customized data analysis on customer side.

\subsection{Requirements and Performance Challenges}
\label{sec:challenges}

The remote measuring use case provides several challenges for development and operations in the mobility cloud. These challenges affect elasticity and resilience properties of the application in practice. 

In regard to elasticity, the first challenge arises from broad range of configuration options in data acquisition campaigns. These campaigns may differ largely in terms of number of recorded vehicle signals, their record frequency, and the number of devices itself. In vehicle homologation, the number of connected devices is rather small, but the number of considered signals is large. Add to this, the data recording frequency in homologation is large and such a campaign is usually triggered in parallel within a few days. This results into single events where large amount of data is pushed into the cloud from the test vehicles. 

In contrast, in fleet observation the number of devices is large, but the number of signals and their recording frequency are small. Due to large number of devices, the accumulated load is large as well, but during fleet observation, we assume that devices connect to the cloud in a rather random and asynchronous profile.

The goal in both usage scenarios is to provide the remote measuring service with acceptable response time for the customer. Hence, remote measuring requires sufficient elasticity to cope with different usage scenarios.

Add to this, the elasticity property of remote measuring determines another service level objective: the cloud costs. While under provisioning causes unacceptable high response times for the users, over provisioning causes high costs, which reduce the revenue of the remote measuring service. Therefore, we search for optimal elasticity property of remote measuring in development and operations.

With respect to resilience, the homologation scenario requires credible data handling to avoid data loss during expensive and elaborate vehicle test rides. Compared with fleet observation, data loss during homologation would cause repetitions of test rides, which can destruct project plans and time to market goals in vehicle development projects. 
In addition, outage of remote measuring during homologation usage would destroy user trust in the application. Due to this, methods to design and test resilience of the applications are of high importance. 

\subsection{Performance Prototype/Demonstrator}
\label{sec:demonstrator}

The remote measuring use case from Figure~\ref{fig:application} was re-imp\-lemented as performance prototype based on the Spring Boot~\footnote{https://spring.io/projects/spring-boot} framework. The components device communication, data provider, data processing, and database were deployed as  containers on an eight node Kubernetes (K8s)~\footnote{https://kubernetes.io/} cluster on bwCloud~\footnote{https://www.bw-cloud.org/}, a state funded academic cloud. The components use the ProtoCom~\footnote{https://sdqweb.ipd.kit.edu/wiki/ProtoCom} library to emulate CPU demands. 

All components and the database were connected through a RabbitMQ~\footnote{https://www.rabbitmq.com/} message broker, which runs on a dedicated node on K8s cluster. The database was deployed as MongoDB~\footnote{https://www.mongodb.com/} container. 
The functionality of the devices was resembled by Gatling~\footnote{https://gatling.io/} load generator. Gatling was used to define load profiles for the system. 
A load profile consists of the number and the ramp up time of the connecting devices, and frequency and size of sent data. Gatling was deployed as container on a dedicated node in K8s.

Several experiments were conducted with different load profiles. Each experiment was triggered as K8s job and the results from Gatling were stored together with monitoring data from Prometheus~\footnote{https://prometheus.io} for following analysis.

\section{Performance Variability of Resource Demand Generation}
\label{sec:performance}

\begin{table}
\scriptsize
\caption{Example calibration run output}
\begin{tabular}{|l|l|l|}
\hline
Time (ms)   & Iterations & Time/Iterations \\ \hline
1,00    & 537389    & 1,86085E-06    \\ \hline
2,00    & 1172345   & 1,70598E-06    \\ \hline
4,00    & 2539921   & 1,57485E-06    \\ \hline
7,97    & 5062500   & 1,57511E-06    \\ \hline
15,85   & 10060004  & 1,5753E-06     \\ \hline
25,73   & 16319999  & 1,57641E-06    \\ \hline
63,44   & 40159726  & 1,57978E-06    \\ \hline
126,68  & 79983883  & 1,58383E-06    \\ \hline
234,10  & 148379031 & 1,57771E-06    \\ \hline
541,72  & 316609902 & 1,71101E-06    \\ \hline
1026,21 & 630079016 & 1,6287E-06     \\ \hline
\end{tabular}
\label{example-calibration-run}
\end{table}
As described previously, to emulate processing of messages in the different services (e.g., the data processing service) each microservice uses ProtoCom. 
ProtoCom requires a low contention calibration phase to determine the input for a particular algorithm (say Fibbonaci number computations) to put load on the CPU for a given time amount (e.g., consume the CPU for 0.2 CPU-seconds). 
The results of the calibration are stored in a model as shown in Table \ref{example-calibration-run} which contains the approximated input parameter associated with their individual execution times. 
Every other resource demand is generated by composing these demands. 
Since the calibration process consumes time (around 20 minutes for \verb|HIGH| accuracy) and the test-bed cluster is homogeneous we initially thought of pre-calibrating ProtoCom and sharing the calibration for all service replicas. This would allow us to execute elasticity experiments and upon the spin-up of new containers, the calibration would not affect the start-up time. 
A precondition for this is, that there is an acceptable variability in CPU time across nodes. Hence, we decide to benchmark the resource demand generation library, namely ProtoCom, to determine how it performs in our cluster. 
\begin{table}[!ht]
\scriptsize
    \centering
    \caption{Kubernetes nodes and their occupancy in number of Pods and average utilization in millicores}
    \begin{tabular}{|l|l|l|l|}
    \hline
        Name & \# Pods & Avg. Millicores & Characteristic App Pods \\ \hline
        minion-01 & 17 & 207.15 & ~ \\ \hline
        minion-02 & 12 & 805.80 & rabbit-broker \\ \hline
        minion-03 & 12 & 152.10 & ~ \\ \hline
        minion-04 & 10 & 176.85 & mongodb \\ \hline
        minion-05 & 8 & 133.30 & ~ \\ \hline
        minion-06 & 7 & 84.50 & ~ \\ \hline
        minion-07 & 11 & 142.20 & demonstrator pods \\ \hline
    \end{tabular}
    \label{tab:number}
\end{table}

\subsection{Environment and Experimentation Setup}

The Kubernetes cluster consists of seven worker nodes of identical flavor \textit{m1.large} with \verb|4vCPUs, 8GB RAM and 12GB| storage. On the worker nodes there are various numbers of container being deployed where some are application-specific and some come from the platform itself. Table \ref{tab:number} summarizes the number of pods per node together with some characteristic application pods. The average millicores determines the average CPU usage of the cluster when no workload is running. 
During the benchmark execution the three services of the application---namely, the Device Communication service, the Data Processing service, and the Data Provider service---are co-located on node \verb|minion-07|.

We define as a \textit{compilation} of the cluster the state after all VMs have been rebooted. We execute the benchmark with two different container QoS classes enabled: \textit{best-effort} (no limit, no guaranteed share) and \textit{guaranteed} (limits are equal to guaranteed share). 
For each class we make five \textit{executions} where in each \textit{execution} five \textit{measurement iterations} follow after an initial five warm-up iterations\footnote{In initial experiments we discovered some warm-up effects affecting the proper calibration of ProtoCom.}. The execution happens on all the seven nodes. A total of 5250 observations determine the performance of ProtoCom in a Kubernetes cluster under two different QoS classes.

To automate the benchmarking process of ProtoCom we make use of the Java Microbenchmark Harness (JMH)\footnote{https://openjdk.java.net/projects/code-tools/jmh/} that facilitates building, running, and analysing (micro-)\hspace{0pt}benchmarks in Java. 
We containerised JMH and use Kubernetes OpenKruise\footnote{https://github.com/openkruise/kruise} to define a \verb|BroadcastJob| that will execute the benchmark on all the nodes in the cluster. We execute five times the benchmark on all nodes. 
In each run the benchmark initially calibrates the ProtoCom library in a \verb|MEDIUM| accuracy setting. There are three different levels of accuracy one can set: \verb|LOW|, \verb|MEDIUM|, and \verb|HIGH|. We chose \verb|MEDIUM| as a compromise between accuracy and experimentation time that showed sufficient stability. 
After the calibration, the benchmark varies the resource demand parameter in three levels 50, 200 and 1000 milliseconds. The selection of the resource demands was motivated from the demands which we inject in the demonstrator application. 

\begin{figure}
  \centering
  \includegraphics[width=0.95\columnwidth]{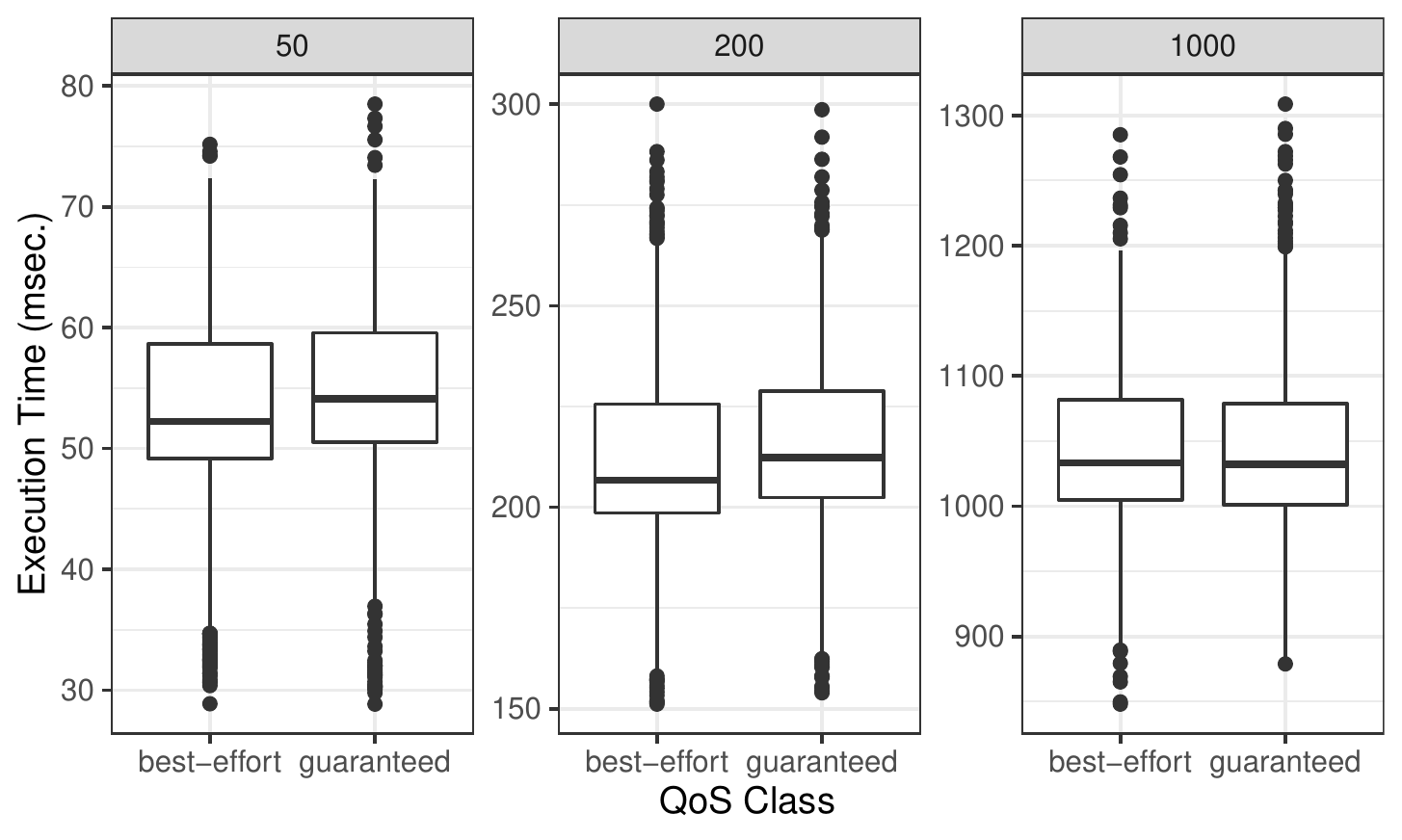}
  \caption{Overall variability of measurements for all instances and executions for the demand parameter set to 50, 200 and 1000 milliseconds in two different Kubernetes QoS classes: \textit{best-effort} and \textit{guaranteed}. 
  }
  \label{fig:overall}
\end{figure}

\subsection{Discussion of Calibration Results}

First, we compare the results for the two used Kubernetes QoS-classes and the three parameter levels.
Figure \ref{fig:overall} summarizes the overall results where the performance is highly variable and observations deviate up to ~40\% in both directions. This happens consistently for the different specified demands. 
The difference between the set container QoS  class is not significant both statistically and practically. The minimal difference stems from the fact that nodes are not highly utilized. Table~\ref{tab:number} shows what additional load (noise) was deployed on the cluster. For the least loaded node, minion-06, variability is low and variability increases for the guaranteed case while the service might get throttled if it has exceeded its limits. For the most loaded node, minion-02, using the guaranteed class shows a positive impact, where for demand 50 the 95th percentile is much closer to the desired target; same applies for demand 1000.
Besides the expected slowdown effects, we also measured several occurrences of speed-ups in our benchmark. 
The box plots show, that the first quartile is matching the desired execution time. Thus 75\% of resource requests are taking more time to complete.
Table \ref{tab:percentiles} shows that for parameters 50 and 200 the 95th percentile and standard deviation is slightly lower when comparing best-effort to guaranteed Kubernetes QoS-class. 
For parameter value 1000 the opposite is true.

\begin{table}[!ht]
\scriptsize
    \centering
    \caption{Execution Time by Parameter and QoS-Class}
    \begin{tabular}{l|l|l|l|l|l}
    \hline
        demand par. & QoS & mean & median & 95th perc. & SD\\ \hline
        50 & best-effort & 52.58131 & 52.248 & 66.5653 & 8.965577 \\ \hline
         & guaranteed & 54.06818 & 54.102 & 65.8925 & 8.115489\\ \hline
        200 & best-effort & 211.0581 & 206.769 & 254.0771 & 24.69515\\ \hline
         & guaranteed & 214.5613 & 212.349 & 253.0308 & 23.41644\\ \hline
        1000 & best-effort & 1042.363 & 1033.147 & 1155.02 & 63.28705\\ \hline
         & guaranteed & 1042.841 & 1032.238 & 1166.254 & 68.88045\\ \hline
    \end{tabular}
    \label{tab:percentiles}
\end{table}

\begin{figure*}
     \centering
     \begin{subfigure}[b]{0.99\columnwidth}
         \centering
         \includegraphics[width=\textwidth]{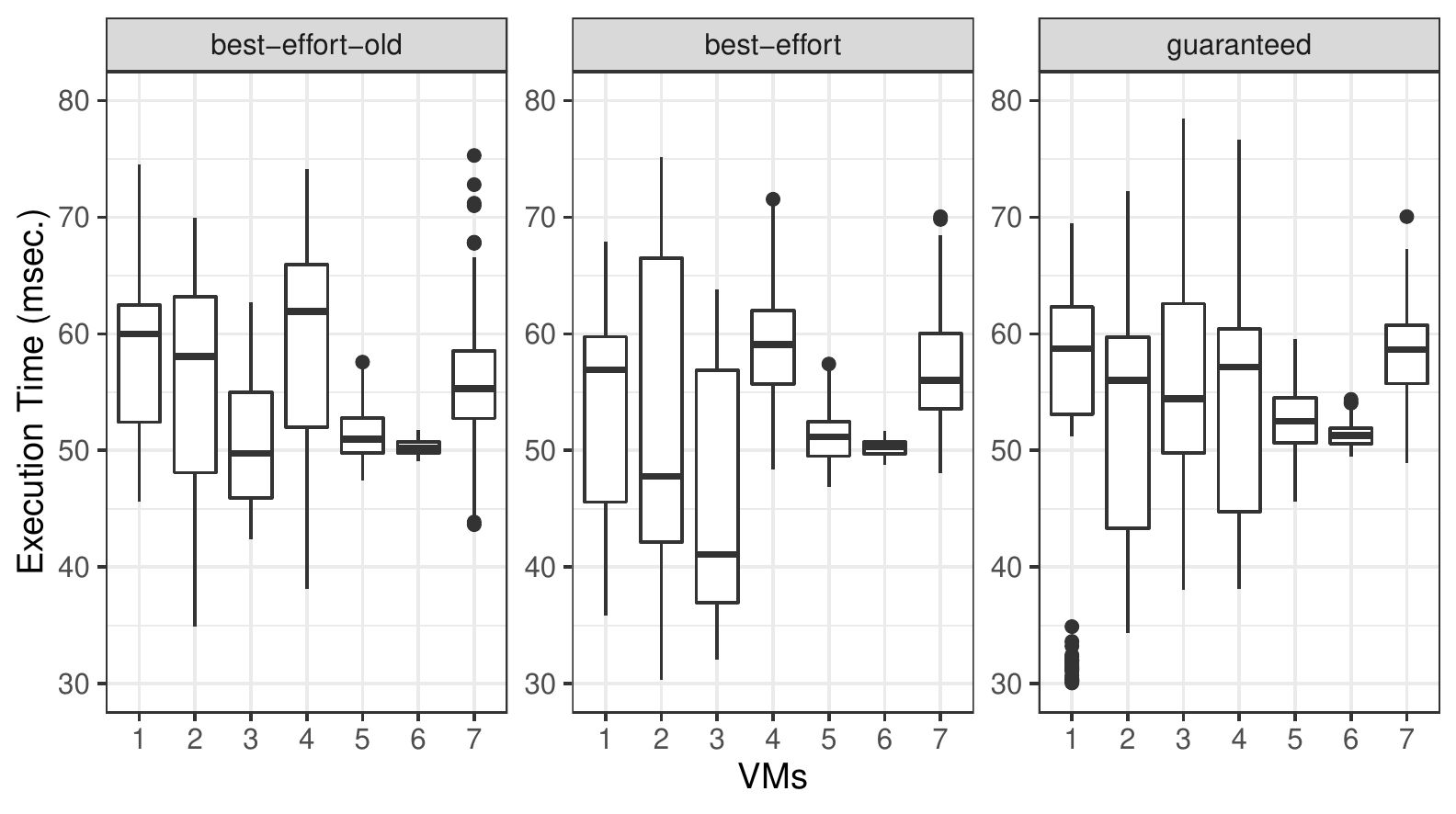}
         \caption{Demand 50}
     \end{subfigure}
     \hfill
     \begin{subfigure}[b]{0.99\columnwidth}
         \centering
         \includegraphics[width=\textwidth]{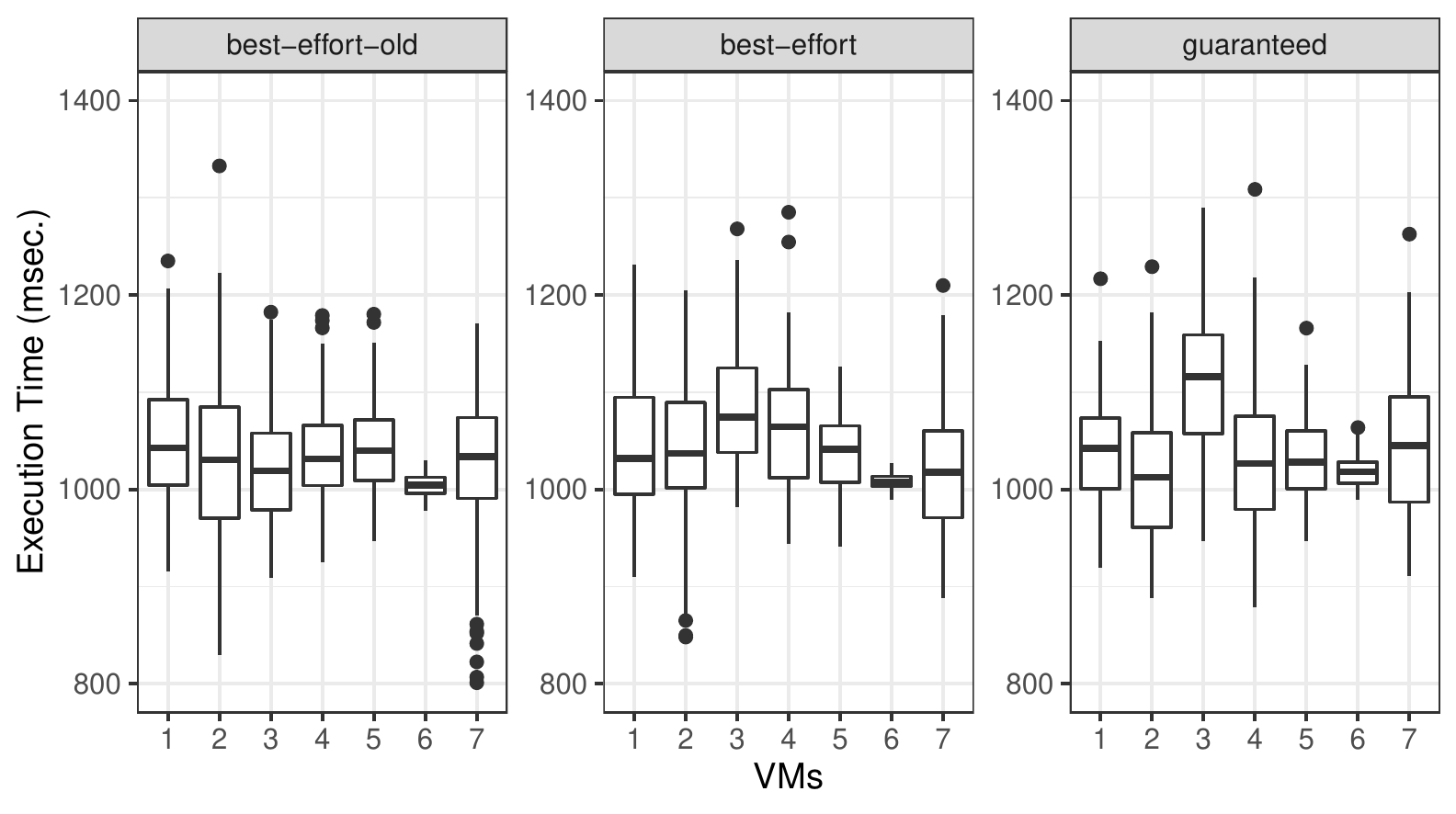}       \caption{Demand 1000}  
     \end{subfigure}
        \caption{Variability of measurements across instances for the demand parameters 50, and 1000 for two different Kubernetes QoS classes: \textit{best-effort} and \textit{guaranteed}. The \textit{best-effort-old} is the execution of the benchmark in a different \textit{cluster compilation} prior to the reboot of VMs for the experiment.}
        \label{fig:pervm}
\end{figure*}

\begin{figure}
     \centering
     \begin{subfigure}[b]{0.23\textwidth}
         \centering
         \includegraphics[width=\textwidth]{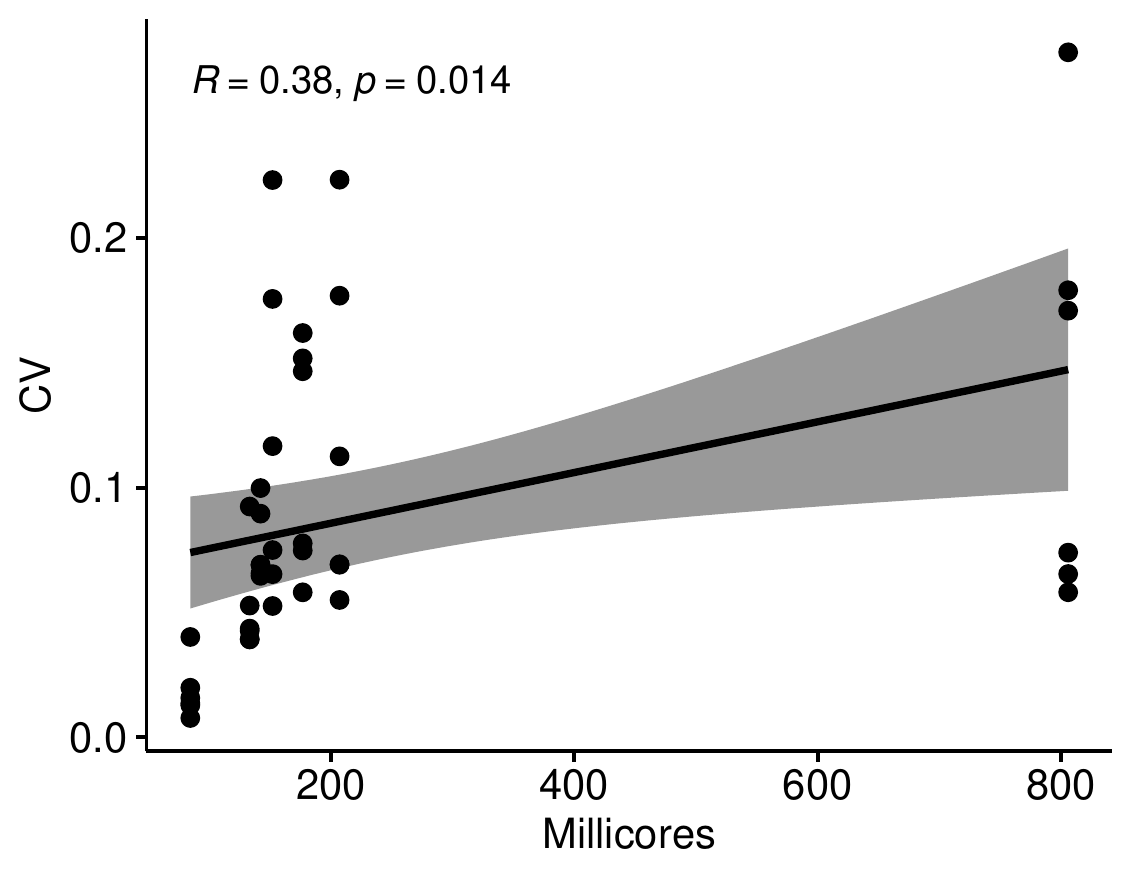}
     \end{subfigure}
     \hfill
     \begin{subfigure}[b]{0.23\textwidth}
         \centering
         \includegraphics[width=\textwidth]{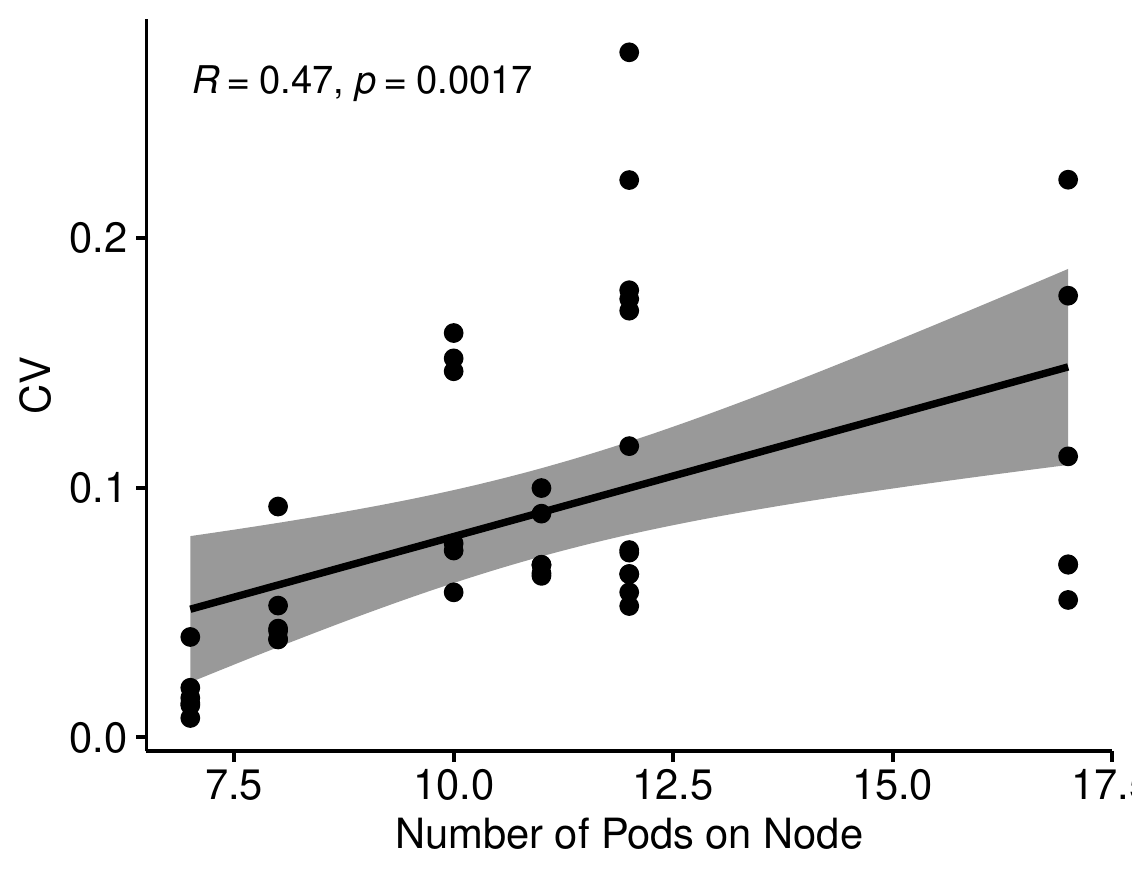} 
     \end{subfigure}
        \caption{The coefficient of variation plotted against the node occupancy for two different measures: average Millicore CPU (left) and the number of pods deployed on a node (right).}
        \label{fig:cor}
\end{figure}

Second, we compare the results of the benchmark across the nodes in the cluster.
We expect to observe differences in the slowdowns due to different placement of VMs in cloud and different level of occupancy of nodes.
As expected, in difference to the QoS class, the variability of the results seems to differ on different nodes. Figure \ref{fig:pervm} depicts the distribution of data points of all iterations on different nodes. The least loaded nodes---in terms of number of pods deployed---experience less variability. 
For estimating the reproducibility of the experiment we added an additional benchmark run (best-effort-old) that was conducted in a different cluster compilation before. 
Results show, though the state of the underlying cloud should have changed, that the variance measured on the nodes is similar.

To analyze the impact of individual node occupancy we calculate the correlation between number of Pods on the node, the CPU utilization in millicores and the sample coefficient of variation (CV) which is the ratio of the standard deviation and the average.
In Kubernetes the utilization is measured in millicores that denotes a thousands of one vCPU,
i.e. a utilization of 207.15 millicores corresponds to 5.18\% for a 4 vCPU node.

Figure \ref{fig:cor} shows a moderate association between both node occupancy, measured in number of pods or millicores CPU, and the relative variability measured through the CV.
To sum up, we show that the intuitive assumption that a higher node occupancy leads to higher variance in execution time during load generation holds.
Moreover, even nodes with moderate average CPU utilization of less than 20\% are affected by slowdowns and varying speeds.
Developers should take that into account when designing performance tests and benchmarks and measure the variability in processing speeds using multiple nodes.
Our simple occupancy metrics can be used as an initial indicator for potential slowdown variations, however, several other factors, like the burstiness of CPU workloads or the I/O usage, could also affect the execution.  

Furthermore, these slowdown effects also impact model-based performance prediction methods, as varying execution time must be taken into account.
While developers could try to model all influencing factors on the nodes in detail, however,
a better solution would be to systematically quantify and model the uncertainty in CPU speeds. 
Based on these information scaling policies and mechanisms can be optimized.


\subsection{Scaling Policies and Mechanisms}
Bondi \cite{Bondi2000} distinguishes four general types of scalability, namely load, space, space-time and structural scalability. In this work we focus only on \textit{load scalability} which is the system's ability 
\textquote[{\cite{Bondi2000}}]{to function gracefully, i.e., without undue delay and without unproductive resource consumption or resource contention at light, moderate, or heavy loads while making good use of available resources}.
Mircoservice.io presents commonly used deployment patterns and distinguishes between the cases of multiple service instances per host and service instance per host, VM or container \cite{richardson2022}.
AWS autoscaling supports a more VM-based scaling approach by creating collections of EC2 instances through scaling a bunch of VMs instead of spinning up containers \cite{awsAutoscaling2022}.
Kubernetes supports scaling on a container level by applying a horizontal pod autoscaling approach, i.e. it assigns more resources by starting additional replicas of a pod (i.e. a container) that is already running for the current workload \cite{kubernetesHpa2021} using the equation:

\begin{equation*}
\scriptsize
\mathrm{replicas}_{\mathrm{desired}} = \mathrm{ceil}\left(\mathrm{replicas}_{\mathrm{current}}\cdot\frac{\mathrm{metric}_{\mathrm{current}}}{\mathrm{metric}_{\mathrm{desired}}} \right). \end{equation*}
\vspace{0.7em}

While this works for any available metric value, commonly an average utilization metric is used.
Pods are placed on existing nodes by the k8s node-scheduler using a two-step filtering and scoring approach\footnote{https://kubernetes.io/docs/concepts/scheduling-eviction/kube-scheduler/}. 

Based on the chosen patterns and technology, software architects might end up with evaluating different kinds of policies.
For the demonstrator application, since it is containerized and deployable in Kubernetes, it is possible to employ both service-based policies and also node-based policies. Here we describe the two different policies and the mechanisms to implement them briefly. 

For the \textbf{service-based} autoscaling policy we rely on the Horizontal Pod Autoscaler (HPA) \cite{kubernetesHpa2021} to define a separate scaling policy for the three different services that constitute the demonstrator application. When fully characterizing the services, various metrics could be used in the configuration of the HPA.  

For the \textbf{node-based} autoscaling policy, we design and implement a controller that replicates Kubernetes nodes and proportionally scales the pods of the three services deployed on the cluster similar to the cluster proportional autoscaler of Kubernetes \cite{cluster-proportional}. A scale out (in) decision occurs whenever the average utilization of available nodes surpasses an upper threshold (falls behind a lower threshold). The number of replicas for each service follows the number of available nodes. For example, if there are two nodes available in the cluster there will be 2 replicas for each of the services.  The initial state constitutes one node whereas there may be up to four nodes where the demonstrator can be allocated. Other nodes are reserved for different services and experimentation tools. 

\begin{algorithm}
\small
\caption{Node-based Autoscaling of Services}\label{alg:cap}
\begin{algorithmic}[1]
\Require $currentAvgUtilization$, $0 \leq upperThreshold \leq 1$, $0 \leq lowerThreshold \leq 1$
\Ensure $nodes$ = $replicasPerDeployment$
\Loop \Comment{Control loop}
\State $currentAvgUtilization$ = $getUtilFromK8s()$
\If{$currentAvgUtilization$ > $upperThreshold$}
    \State makeOneAdditionalNodeAvailable()
    \State scaleOutAllDeploymentsByOne()
\EndIf
\If{$currentAvgUtilization$ < $lowerThreshold$} 
    \State makeOneNodeUnavailable()
    \State scaleInAllDeploymentsByOne()
\EndIf
\State sleep()
\EndLoop
\end{algorithmic}
\end{algorithm}

The two different policies lead to different \textit{cluster compilations} over time. In the next section we explore the scalability boundaries of the demonstrator when assuming a node-based autoscaling policy. This allows us to design meaningful scenarios for the elasticity experiments and later evaluate various autoscaling policies including the described ones.

\subsection{Scalability Assessment as a Prerequisite of Designing Elasticity Scenarios} 

Several proposals exist in literature to asses the scalability of applications, e.g., \cite{DBLP:conf/wosp/HenningH21}. The main purpose of assessing the scalability in our case is to estimate the processing limits of the application and use this information for designing elasticity experiments. To determine whether a configuration can handle a certain load (in terms of devices) we observe the performance and the utilization of the system. We follow a binary search procedure \cite{DBLP:conf/usenix/ShivamMCSB08} to reach faster the upper bound on the number of devices that a particular configuration can handle without violating a given performance service level objective (SLO). 
For the remote measuring use case, we define the target SLO to be one second response time for the 95th percentile during a constant load of active devices. Since we lack data for workload characterization for the reference application, we generate synthetic load using Gatling that stresses the application in two configurations: the initial configuration where the demonstrator is deployed on one node and the final configuration where the demonstrator is deployed on four nodes. 

\begin{figure}
     \centering
     \includegraphics[width=0.99\columnwidth]{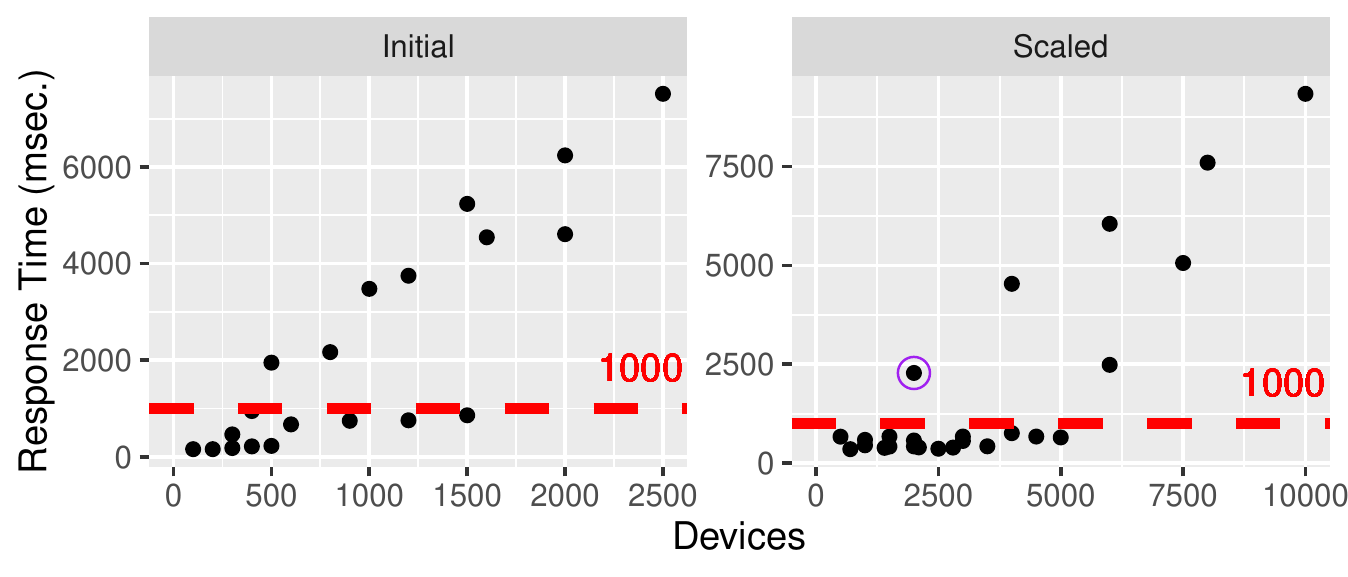}
     
    \caption{The 95th percentile of response time of the \textit{initial} and \textit{scaled} cluster compilation for synthetic generated load assuming that the application scales using a node-based autoscaling policy.}
    \label{fig:scalability}
\end{figure}

As Figure \ref{fig:scalability} depicts, the demonstrator application is able to scale with additional resources i.e., when such resources are provided by means of VM replication and scaling proportionally the corresponding pods.  One data point in the plot shows the 95th percentile of response time for requests that have been generated by the active devices given on the x-axis. 
The red dashed line shows the SLO boundary of 1000ms that should not be violated.
The synthetic workload influences response times through the number of concurrent users (devices), how fast they ramp-up, and the sleep value. The combination of the time to ramp up (5 seconds), the frequency of sending data (60 seconds) and the number of devices together with the intrinsic randomness of the workload generation tool and setup leads to different distributions of inter-arrival times for requests. As the highlighted data point in the scaled part depicts that spawning 2000 devices in 5 seconds in one case leads to the violation of the SLOs while the configuration is able to sustain higher number of devices---and up to a maximum of 5000 in one observation---when the devices are increased gradually. 

Initial experiments allow us to estimate the performance boundaries of the application when assuming a \textbf{node-based} scaling policy for the \textit{initial} and the most \textit{scaled} configuration that may occur. However, when employing a \textbf{service-based} policy, although the initial configuration might be assumed similar, the most scaled configuration is not easily derived upfront. However, the approximated processing capabilities are used for designing scenarios for elasticity experiments in which both alternatives for autoscaling policies could be evaluated.
\section{Conclusions and Future Work} \label{sec:conclusions}
Existing reference applications for experimenting and research are not representative for data-intensive containerized cloud applications. They are serving more traditional use cases of human-centered request-reply communication without a continuous data processing pipeline and without using asynchronous messaging for the communication. 

In attempt to fill this gap, in this work we present a reference use case coupled with the initial architecture design and with the engineering challenges for elasticity and resilience. To make the reference application suitable for research of elasticity and resilience mechanisms through increasing the predictability in performance, we have experimented with the CPU load generation tool ProtoCom. Results show a high variability in execution time when generating load for a given time demand. We observe a moderate association between node occupancy and the relative variability. In addition, we sketch how the application could be scaled through two different autoscaling policies and investigate the scalability of the solution assuming a node-based approach.

In the future we plan to investigate further on the performance influences of using different QoS classes for containers. Moreover, we will perform additional experiments using container quotas and CPU throttling. 
Consolidating and publishing the reference application including the evaluation of elasticity and resilience scenarios is another item for future work. In parallel we have started to construct a performance model of the application that will allow the evaluation of architecture alternatives.
Moreover, we created initial simulation models to emulate the QoS-based shared CPU scheduling regime used in Kubernetes and alike.
In the long run, we want to optimize scalability and resilience for dynamic cloud applications by incorporating models of uncertain execution environments~\cite{9196340} into simulation-based prediction techniques.

\section*{Acknowledgments}
This paper was partly funded by the Federal Ministry of Education and Research under grant number 01IS18069A. 
For computational resources we acknowledge the bwCloud (https://www.bw-cloud.org), funded by the Ministry of Science, Research and Arts Ba\-den-Würt\-tem\-berg (Mi\-nis\-te\-ri\-um für Wis\-sen\-schaft, For\-schung und Kunst Ba\-den-Würt\-tem\-berg).

\bibliography{references}{}

\begin{thebibliography}{10}

\bibitem{acmeAir}
{ACME Air: Acme Air Sample and Benchmark }.
\newblock Online: \url{https://github.com/acmeair/acmeair}.
\newblock [Online; 2022-01-25].

\bibitem{cluster-proportional}
{Cluster Proportional Autoscaler}.
\newblock Online:
  \url{https://github.com/kubernetes-sigs/cluster-proportional-autoscaler}.
\newblock [Online; 2022-01-25].

\bibitem{sockshop}
Sockshop microservice demo application.
\newblock Online: https://microservices-demo.github.io.

\bibitem{springpetclinic}
{Spring PetClinic}.
\newblock Online: \url{https://github.com/spring-project:spring-petclinic.}
\newblock [Online; 2022-01-25].

\bibitem{kubernetesHpa2021}
{Kubernetes Horizontal pod autoscaling}.
\newblock
  \url{https://kubernetes.io/docs/tasks/run-application/horizontal-pod-autoscale/},
  2021.
\newblock [Online; 2021-12-09].

\bibitem{awsAutoscaling2022}
{Amazon EC2 Auto scaling}.
\newblock
  \url{https://docs.aws.amazon.com/autoscaling/ec2/userguide/what-is-amazon-ec2-auto-scaling.html},
  2022.
\newblock [Online; 2022-01-25].

\bibitem{10.1007/978-3-540-69814-2_7}
Steffen Becker, Tobias Dencker, and Jens Happe.
\newblock Model-driven generation of performance prototypes.
\newblock In Samuel Kounev, Ian Gorton, and Kai Sachs, editors, {\em
  Performance Evaluation: Metrics, Models and Benchmarks}, pages 79--98,
  Berlin, Heidelberg, 2008. Springer Berlin Heidelberg.

\bibitem{Bondi2000}
Andr{\'{e}}~B. Bondi.
\newblock Characteristics of scalability and their impact on performance.
\newblock In {\em Proceedings of the second international workshop on Software
  and performance - {WOSP} {\textquotesingle}00}. {ACM} Press, 2000.

\bibitem{bosch21}
Bosch.
\newblock {Mobility Cloud}.
\newblock
  \url{https://www.bosch-mobility-solutions.com/media/global/products-and-services/mobility-services/plcs/mobility-cloud/21-08-02_bosc_21028-06_mobilitycloud_onepager-rgb_en.pdf},
  2021.
\newblock [Online; 2021-12-16].

\bibitem{10.5555/2636651}
Marco Di~Natale, Haibo Zeng, Paolo Giusto, and Arkadeb Ghosal.
\newblock {\em Understanding and Using the Controller Area Network
  Communication Protocol: Theory and Practice}.
\newblock Springer Publishing Company, Incorporated, 2014.

\bibitem{DBLP:conf/wosp/HenningH21}
S{\"{o}}ren Henning and Wilhelm Hasselbring.
\newblock How to measure scalability of distributed stream processing engines?
\newblock In Johann Bourcier, Zhen Ming~(Jack) Jiang, Cor{-}Paul Bezemer,
  Vittorio Cortellessa, Daniele~Di Pompeo, and Ana~Lucia Varbanescu, editors,
  {\em {ICPE} '21: {ACM/SPEC} International Conference on Performance
  Engineering, Virtual Event, France, April 19-21, 2021, Companion Volume},
  pages 85--88. {ACM}, 2021.

\bibitem{laaberMicrobenchmarking}
Christoph Laaber, Joel Scheuner, and Philipp Leitner.
\newblock Software microbenchmarking in the cloud. how bad is it really?
\newblock {\em Empirical Softw. Engg.}, 24(4):2469–2508, aug 2019.

\bibitem{lehrig2018cloudstore}
Sebastian Lehrig, Richard Sanders, Gunnar Brataas, Mariano Cecowski, Simon
  Ivan{\v{s}}ek, and Jure Polutnik.
\newblock Cloudstore—towards scalability, elasticity, and efficiency
  benchmarking and analysis in cloud computing.
\newblock {\em Future Generation Computer Systems}, 78:115--126, 2018.

\bibitem{lehrig2011performance}
Sebastian Lehrig and Thomas Zolynski.
\newblock Performance prototyping with protocom in a virtualised environment: A
  case study.
\newblock {\em Proceedings to Palladio Days}, pages 17--18, 2011.

\bibitem{Lutz2017}
Albert Lutz, Bernhard Schick, Henning Holzmann, Michael Kochem, Harald
  Meyer-Tuve, Olav Lange, Yiqin Mao, and Guido Tosolin.
\newblock Simulation methods supporting homologation of electronic stability
  control in vehicle variants.
\newblock {\em Vehicle System Dynamics}, 55(10):1432--1497, 2017.

\bibitem{nikounia2015hypervisor}
Seyed~Hossein Nikounia and Siamak Mohammadi.
\newblock Hypervisor and neighbors’ noise: Performance degradation in
  virtualized environments.
\newblock {\em IEEE Transactions on Services Computing}, 11(5):757--767, 2015.

\bibitem{richardson2022}
Chris Richardson.
\newblock {microservices.io deployment patterns}.
\newblock
  \url{https://microservices.io/microservices/news/2015/03/15/deployment-patterns.html},
  2021.
\newblock [Online; 2022-01-25].

\bibitem{rosen2014linux}
Rami Rosen.
\newblock Linux containers and the future cloud.
\newblock {\em Linux J}, 240(4):86--95, 2014.

\bibitem{9196340}
Max Scheerer, Martina Rapp, and Ralf Reussner.
\newblock Design-time validation of runtime reconfiguration strategies: An
  environmental-driven approach.
\newblock In {\em 2020 IEEE International Conference on Autonomic Computing and
  Self-Organizing Systems (ACSOS)}, pages 75--81, 2020.

\bibitem{DBLP:conf/usenix/ShivamMCSB08}
Piyush Shivam, Varun Marupadi, Jeffrey~S. Chase, Thileepan Subramaniam, and
  Shivnath Babu.
\newblock Cutting corners: Workbench automation for server benchmarking.
\newblock In Rebecca Isaacs and Yuanyuan Zhou, editors, {\em 2008 {USENIX}
  Annual Technical Conference, Boston, MA, USA, June 22-27, 2008. Proceedings},
  pages 241--254. {USENIX} Association, 2008.

\bibitem{von2018teastore}
J{\'o}akim von Kistowski, Simon Eismann, Norbert Schmitt, Andr{\'e} Bauer,
  Johannes Grohmann, and Samuel Kounev.
\newblock Teastore: A micro-service reference application for benchmarking,
  modeling and resource management research.
\newblock In {\em 2018 IEEE 26th International Symposium on Modeling, Analysis,
  and Simulation of Computer and Telecommunication Systems (MASCOTS)}, pages
  223--236. IEEE, 2018.

\bibitem{DBLP:journals/tse/ZhouPXSJLD21}
Xiang Zhou, Xin Peng, Tao Xie, Jun Sun, Chao Ji, Wenhai Li, and Dan Ding.
\newblock Fault analysis and debugging of microservice systems: Industrial
  survey, benchmark system, and empirical study.
\newblock {\em {IEEE} Trans. Software Eng.}, 47(2):243--260, 2021.

\end{thebibliography}
\bibliographystyle{plain}

\end{document}